\documentclass[aps,twocolumn, pra,showpacs]{revtex4-1}

\usepackage{epsfig,amssymb,amsmath,graphicx,color}

\begin{document}

\title{Scattering distributions in the presence of measurement backaction}
\author{James S. Douglas}
\affiliation{Institut de Ciencies Fotoniques, 08860 Castelldefels, Barcelona, Spain}
\author{Keith Burnett}
\affiliation{University of Sheffield, Western Bank, Sheffield S10 2TN, United Kingdom}
\begin{abstract}
Scattering probe particles from a quantum system can provide experimental access to information about the system's state. However, measurement backaction and momentum transfer during scattering changes the state of the system, potentially destroying the state of the system we wish to probe. Here we investigate how to probe the system's initial state even in the presence of backaction and momentum transfer. We show that summing the scattering distributions of an ensemble of measurements reveals the initial state scattering pattern even when each of the individual measurements completely destroys the initial state. This procedure is effective provided the scattering takes place on a timescale that is short compared with the free evolution of the system.
\end{abstract}
\maketitle

\section{Introduction}

Scattering of particles from matter is a fundamental method used to reveal a system's properties. Rutherford's revolutionary discovery of the atomic nucleus involved scattering alpha particles from gold foil \cite{Rutherford1911a}, while scattering of x-ray photons has led to numerous discoveries about crystalline structure, such as the structure of DNA \cite{Franklin1953a}. The range of phenomena investigated by scattering is broad, where for example neutron scattering has been used to probe the nature of exotic forms of matter such as superfluid helium and high temperature superconductors \cite{Griffin1993a}, and scattering of optical photons has been used to probe correlations within Bose-Einstein condensates \cite{Stamper-Kurn1999a}.
In all these cases probe particles scatter from the matter because of density or charge fluctuations in the system or by creating or destroying excitations in the system. Measuring the scattering pattern then yields valuable information about the system's spatial structure and excitations, and furthermore interference between different scattering channels reveals information about correlations within the system.

Typically in experiments and theoretical discussions involving scattering, the effect of the probe on the system is assumed to be weak and the scattering pattern is calculated in the Born approximation from the initial system state \cite{VanHove1954}. However during each scattering event energy and momentum may be transferred from the probe to the system. Furthermore the act of measuring the probe's scattering pattern may produce quantum backaction on the state of the system \cite{Hatridge2013a,Javanainen2013a}. These effects can combine to significantly alter the state of the system, indeed for two particles, scattering of light leads to localization of relative position while the particles' momentum states become entangled \cite{Rau2003a,Cable2005a,Dunningham2012a}. This complicates the use of scattering to probe the state of a system, as after a small number of scattering events the system may be in a completely different state from which it started. 

Here we model the evolution of manybody states caused by consecutive scattering and measurement, and see how quantum backaction changes the scattering distribution for multiple particle systems. As we have previously shown, scattering leads to localization of the system particles in position space and, in some regimes, superpositions of localized particle configurations \cite{Douglas2012a}. 
Here we show that the average scattering pattern for an ensemble of experiments gives the pattern predicted for the initial state even if this initial state is completely destroyed in each individual scattering experiment. This is possible when the scattering measurements all occur on a timescale short compared with the free evolution of the system. Experimental access to the initial state of the system is hence possible even for systems that are easily perturbed. This is an important result that we expect can be utilized, for example, in the field of quantum simulation, where experiments using cold atoms are used to simulate many-body physics \cite{Lewenstein2007a,Bloch2008a}. The outcome of each cold atom experiment can be measured by scattering light or matter from the system \cite{McKay2011a,Mekhov2012a,Sanders2010a,Gadway2011a,Douglas2011a,Douglas2010a,Douglas2011c} and then the state of the system can be reproduced by combining results from multiple runs of the experiment.

\section{Consecutive scattering from manybody states}

The model of consecutive scattering that we use was presented in Ref.~\cite{Douglas2012a} and we reproduce the relevant results here for completeness. We consider coherent scattering of a probe beam from a system of identical particles, where the probe particles are assumed to be in approximately plain wave form with initial wave-vector $\mathbf{k}_i$. Through scattering the probe can acquire a new wave-vector $\mathbf{k}_f$, which is then associated with a momentum transfer to the system of $\hbar\mathbf{k} =\hbar(\mathbf{k}_i - \mathbf{k}_f)$. For a weak probe-system interaction the probability of multiple scattering is low and we can then assume this momentum transfer occurs in a single scattering event.
For an $N$-particle state $|\psi\rangle = \int d\mathbf{R} \psi(\mathbf{R})|\mathbf{R}\rangle$, where  $\mathbf{R}= \{\mathbf{r}_1,\ldots,\mathbf{r}_N\}$ gives the coordinates of the $N$ particles, the scattering interaction results in a combined probe-system state of
\begin{equation}
|\psi\rangle\otimes|\mathbf{k}_i\rangle \longrightarrow \int d\mathbf{R} d\mathbf{k}_f  S(\mathbf{k},\mathbf{R})\psi(\mathbf{R})|\mathbf{R}\rangle\otimes|\mathbf{k}_f\rangle
\label{eq:total_transform}
\end{equation}
where
\begin{equation}
S(\mathbf{k},\mathbf{R}) =\left\{\begin{array}{ccc}
 \frac{g}{\sqrt{4 \pi}}\sum_{j=1}^N e^{i \mathbf{k}\cdot\mathbf{r}_j} & \mbox{for} & \mathbf{k}_f \neq  \mathbf{k}_i\\
A(\mathbf{R}) &  \mbox{for} & \mathbf{k}_f =  \mathbf{k}_i \end{array}\right.
\end{equation}
This state is superposition of all possible single scattering events along with the possibility that the probe did not scatter and remains in $|\mathbf{k}_i\rangle$ \cite{Douglas2012a}. The factor $g$ is given by the strength of the probe-system interaction and for simplicity we assume $g$ is independent of $\mathbf{k}_f$, which is the case when energy conservation restricts scattering so that $|\mathbf{k}_f|\sim|\mathbf{k}_i|$ \cite{VanHove1954}. This occurs when the probe particles are much lighter than the system particles, as in the case of light scattering, or can occur as we see below in the case of cold atoms in optical lattices where the system particles are confined to energy bands. We further assume the probe-system interaction is isotropic, as is the case for $s$-wave scattering in cold atoms. The treatment can be generalized to take into account anisotropic scattering, such as the dipole pattern in light scattering, but this does not affect the qualitative behavior discussed below.

Measurement of the final probe wave-vector, for example by imaging in the far field, causes a backaction on the system state collapsing this superposition $|\psi\rangle \longrightarrow \int d\mathbf{R}S(\mathbf{k},\mathbf{R})\psi(\mathbf{R})|\mathbf{R}\rangle$. If a probe particle is observed to be scattered with wave-vector $\mathbf{k}_f$ then the new system state is a superposition of each system particle having received momentum $\hbar\mathbf{k}$ from the probe, given the new state a broader relative momentum distribution. The superposition occurs as detecting the probe's final wave-vector reveals no information about which particle scattered it. As the system's state broadens in relative momentum space it localizes in relative position space \cite{Rau2003a}.
However, because the scattering probability distribution is completely determined by the relative positions of pairs of system particles, the backaction of the measurement on the state will then preserve superpositions of states that have the same set of relative position vectors. Measuring the scattering distribution can then lead to spatial superpositions of the system particles as we describe in Ref.~\cite{Douglas2012a}.

Due to the small value of $g$ it is often the case that the probe particles do no scatter from the system. Naively one would assume that this does not affect the system, but the measurement of a non-scattering event also reveals information about the state and leads to measurement backaction \cite{Rau2003a}. This is because some system configurations scatter less than others and observing non-scattering projects the state toward these configurations. The non-scattering amplitude $A(\mathbf{R})$ is fixed by the condition that probe particles must either be scattered or non-scattered, that is $1 =A(\mathbf{R})^2 + \int_{\mathbf{k}\neq 0} d\mathbf{k}|S(\mathbf{k},\mathbf{R})|^2\equiv  \int  d\mathbf{k}|S(\mathbf{k},\mathbf{R})|^2$.

Together consecutive scattering and non-scattering events lead to a dynamic evolution of the many-body state. Neglecting, for the moment, the free evolution of the system between each scattering event, the net result of the scattering evolution is that each scattering measurement provides a partial measurement of the relative positions of the system particles. In the limit of an infinite number of scattering events the relative positions of the system particles are determined exactly, at least in principle, and the resulting state is an eigenstate of relative position, 
%(or superposition of eigenstates corresponding to the same scattering pattern)
such as the completely localized state $|\mathbf{R}\rangle$. The probability of the state ending in $|\mathbf{R}\rangle$ is given by the initial state wavefunction $|\psi(\mathbf{R})|^2$ and the scattering pattern then observed is $|S(\mathbf{k},\mathbf{R})|^2$. If we then do multiple repetitions of the experiment and image the scattering pattern for each, we will get an ensemble of scattering patterings corresponding to various $|\mathbf{R}\rangle$ with probability $|\psi(\mathbf{R})|^2$. Summing the scattering patterns over the ensemble then gives the same result as the scattering pattern for the initial state (without backaction)
$\int d\mathbf{R}\left|S(\mathbf{k},\mathbf{R})\psi(\mathbf{R})\right|^2$. 

More generally for a finite number of scattering events,  $m$, the state becomes $\int d\mathbf{R} S(\mathbf{k}_1,\mathbf{R})\ldots S(\mathbf{k}_m,\mathbf{R})\psi(\mathbf{R})|\mathbf{R}\rangle$. Then in an ensemble of measurements with $m$ events each, any scattering event we choose, say $j$, has a probability distribution given by $\int d\mathbf{k}_1 \ldots d\mathbf{k}_{j-1}d\mathbf{k}_{j+1}  \ldots d\mathbf{k}_m|S(\mathbf{k}_1,\mathbf{R})\ldots S(\mathbf{k}_m,\mathbf{R})\psi(\mathbf{R})|^2 = \int d\mathbf{R} |S(\mathbf{k}_j,\mathbf{R})\psi(\mathbf{R})|^2$, that is the initial state pattern.
Combining the scattering patterns from an ensemble of measurements then gives an estimate of the initial state scattering pattern with a sampling error that scales with $1/\sqrt{n}$ where $n$ is the number of experimental runs. This occurs despite the fact that scattering leads the initial state to be destroyed in each individual run.

\section{Scattering and free evolution in the Bose-Hubbard model} 

The analysis above assumes the system does not evolve while the scattering measurements are taking place. To examine how the system's free evolution affects the measurement we now examine a specific example that displays the key characteristics of the dynamic scattering process. We consider scattering from a system of Bosons on a one-dimensional lattice described by the Bose-Hubbard model. This can be achieved in experiment by matter-wave scattering from a one-dimensional lattice of bosonic atoms in an optical lattice \cite{Jaksch1998b,Sanders2010a,Gadway2011a}. We take the lattice to have $M$ sites and to be oriented along the $y$-axis, while the initial wave-vector of the probe is in the $x$-direction. For simplicity we only consider scattering within the $xy$-plane, where the full three dimensional scattering is a straightforward generalization.

At low temperature the atoms all reside in the lowest band of the lattice and the energy of the probe can be arranged so that excitation of the atoms in the lattice to higher bands is negligible \cite{Sanders2010a}. The state of the system can then be described in terms of  lowest band Wannier functions $w(\mathbf{r}-\mathbf{r}_j)$, where $r_j$ is the position of the $j$th lattice site \cite{Kittel}.
%The atomic field operator can then be reduced to an expansion in terms of lowest band Wannier functions $\hat{\Psi}(\mathbf{r}) = \sum_j w(\mathbf{r}-\mathbf{r}_j) \hat{a}_j$, where $r_j$ is the position of the $j$th lattice site \cite{Kittel}.
 When this condition is not met we expect to get similar results by taking into account the contributions from higher bands \cite{Douglas2011a}.
 We assume that the lattice potential is strong enough that the overlap between neighboring Wannier functions is negligible. The state of the system can be expressed in terms of a number basis $|\mathbf{n}_u\rangle \equiv|\{n^{u}_j,j=1,\ldots,M\}\rangle$, where $n_j^{u}$ is the number of atoms at site $j$ and $u$ uniquely identifies each basis state. After the $m$th scattering event we expand the state as $|\Psi_m\rangle = \sum_u \psi_u^{m}|\mathbf{n}_u\rangle$. Scattering from this state then occurs at angle $\theta$ to the $x$-axis with probability
\begin{equation}
P_m(\theta) = \frac{g^2}{2 \pi}\sum_u \left|I(\theta) \psi_u^{m}\sum_j e^{i \mathbf{r}_j \cdot \mathbf{k}(\theta)} n_j^{u}\right|^2.
\label{eq:ang_prob_density}
\end{equation}
where $I(\theta) = \int d\mathbf{r} e^{i \mathbf{k}(\theta) \cdot\mathbf{r}} |w(r)|^2$ and $\mathbf{k}(\theta)=k_0(1-\cos\theta,-\sin\theta)$.

Following a detection at $\theta$ the many-body state is projected into the new state
\begin{equation}
|\Psi_{m+1}\rangle = \frac{1}{\sqrt{\mathcal{N}}}\sum_u \psi_u^{m}\sum_j e^{i \mathbf{r}_j \cdot \mathbf{k}(\theta)} n_j^{u}|\mathbf{n}_u\rangle,
\label{eq:theta_projection}
\end{equation}
where $\mathcal{N}$ normalizes the state.
We note that the number basis states are eigenstates of this projection, and the scattering process will preserve any state that begins in a basis state. Moreover, some of the basis states produce the same scattering pattern as the relative positions of pairs of atoms in the lattice are the same, for example in the $N=M=3$ case, $|2 0 1\rangle$ and $|1 0 2\rangle$ result in the same light scattering. Superposition of these states are partially preserved by the projection, in that the weight of each state in the superposition remains the same after scattering but the phase relationship is changed. 
%For initial states that are the superposition of all the basis states, the projection gives higher weight to the basis states where the relative position is zero for each pair of particles, that is the state where $n_s = N$, for some lattice site $s$, and all other sites have zero occupancy. These states have the highest probability of scattering at any angle and detecting a scattering event makes it more probable that the atomic system is in one of these states. A sequence of scattering events with $\theta\neq 0$ then leads to a superposition of these states in the limit $m\rightarrow \infty$. 

Alternatively detection events where the probe is not scattered occur with probability
\begin{equation}
P_m^{NS} = \sum_u |\psi_u^{m} A_u|^2
\label{eq:prob_non_scat}
\end{equation}
and
\begin{equation}
A_u = \sqrt{1-\frac{g^2}{2\pi} \int_{-\pi}^\pi d\theta \left|I(\theta)\sum_j e^{i \mathbf{r}_j\cdot\mathbf{k}(\theta)} n_j^{u}\right|^2}.
\end{equation}
Detecting a non-scattering event projects the state into the new state
\begin{equation}
|\Psi_{m+1}\rangle = \frac{1}{\sqrt{\mathcal{N}'}}\sum_u \psi_u^{m}A_u|\mathbf{n}_u\rangle,
\label{eq:non_scat_projection}
\end{equation}
which favors states with a lower probability of scattering.

The dynamic scattering process can now be simulated using a quantum jump procedure \cite{Dalibard1992a,Dum4879a}. Taking the initial state, we calculate the probability distributions for scattering and non-scattering. A pseudo-random number is then used to determine if scattering occurs and in which direction. If it does then the projection in Eq.~(\ref{eq:theta_projection}) is applied, if it does not the non-scattering projection in Eq.~(\ref{eq:non_scat_projection}) is applied. In either case the many-body state is then normalized and evolves according to Bose-Hubbard Hamiltonian before becoming the input state and the process repeats. For simplicity we assume the scattering events are evenly spaced by time $dt$.

%%%%%%%%%%%%%%%%%%%%%%% FIGURE %%%%%%%%%%%%%%%%%%%%%%%%%%%%%%%%%%%%%%%%%%%%%%%%%%%%%%%%%%%%%%%%%%%%%%%%%%%%%%%%%%%%%%%%%%%%%%
%%%%%%%%%%%%%%%%%%%%%%%%%%%%%%%%%%%%%%%%%%%%%%%%%%%%%%%%%%%%%%%%%%%%%%%%%%%%%%%%%%%%%%%%%%%%%%%%%%%%%%%%%%%%%%%%%%%%%%%%%%%%%

\begin{figure}
\centering
\includegraphics{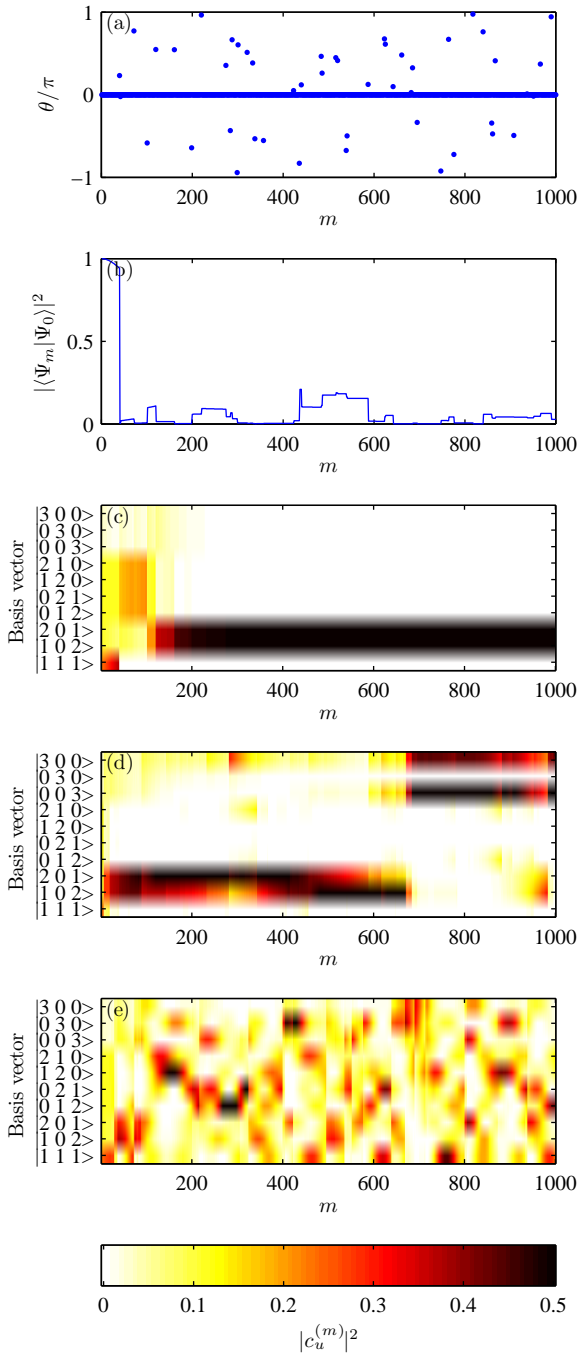}
\caption{Development of an atomic state caused by 1000 scattering events for a lattice with three sites and three atoms. (a) Detected events. (b) Overlap of the many-body state with the initial ground state $|\Psi_0\rangle$. (c)-(e) Modulus squared of the basis coefficients of the state $|\Psi_m\rangle$ for (c) $dt = 0$ (b) $dt = 0.001 J/\hbar$ and (c) $dt = 0.01 J/\hbar$. Parameters used are $U/J = 0.05$, $g=0.1$ and $k_0 = \pi/a$, where $a$ is the lattice site separation and $U$ and $J$ are the parameters of the Bose-Hubbard model.}
\label{fig:develop_three_site}
%figure produced by Dynamic_light_scatter_with_non_scatter_sf_only_after_scat.m
\end{figure}

%%%%%%%%%%%%%%%%%%%%%%% FIGURE %%%%%%%%%%%%%%%%%%%%%%%%%%%%%%%%%%%%%%%%%%%%%%%%%%%%%%%%%%%%%%%%%%%%%%%%%%%%%%%%%%%%%%%%%%%%%%
%%%%%%%%%%%%%%%%%%%%%%%%%%%%%%%%%%%%%%%%%%%%%%%%%%%%%%%%%%%%%%%%%%%%%%%%%%%%%%%%%%%%%%%%%%%%%%%%%%%%%%%%%%%%%%%%%%%%%%%%%%%%%

As an example of the dynamic process we look at the simple case of a three site lattice containing three atoms. In this case there are only ten basis states making it straightforward to track the development of the many-body state. 
In Figure \ref{fig:develop_three_site} we show realizations of the dynamic scattering process for the three site lattice in the superfluid regime of the Bose-Hubbard model. For this example we have set the coupling constant $g=0.1$ and as a result the vast majority of detection events result from non-scattering, shown as $\theta = 0$ in Figure \ref{fig:develop_three_site}(a). In Figure \ref{fig:develop_three_site}(c) we show the development of the many-body state for the case of no system evolution between each scattering event, $dt = 0$. In this example the many-body state progresses toward a superposition of the states $|2 0 1\rangle$ and $|1 0 2\rangle$, two states which produce the same scattering pattern. As discussed above, continued scattering from this end superposition does not change the constituent basis vectors but does change the phase of the superposition. In Figure \ref{fig:develop_three_site}(b) we see that detection at non-zero angle quickly reduces the overlap of the many-body state with the original ground state. We see that the overlap makes quantum jumps when a scattering event occurs and gradually evolves due to non-scattering events.

Figure \ref{fig:develop_three_site}(d) and (e) demonstrate the effect of system evolution between each scattering event. For the example evolution in Figure \ref{fig:develop_three_site}(d) with $dt = 0.001 J/\hbar$ the system progresses toward a superposition like that in Figure \ref{fig:develop_three_site}(b), however after some time the system evolution causes the system to jump to another set of states. In Figure \ref{fig:develop_three_site}(e) with $dt = 0.01 J/\hbar$ the evolution is completely different with the system exploring a range of basis states over the simulation. This is detrimental to determining the initial state scattering pattern and, as we will see below, to do so we must limit the time over which scattering occurs to less than the natural system evolution time.

\section{Scattering distributions from ensemble measurements}

By repeated simulation of the dynamic scattering process with $dt = 0$, we find that the simulations all settle into a final state after a small number of scattering events, as in Figure \ref{fig:develop_three_site}(c). The end states are always superpositions of eigenstates of the scattering projection that produce the same scattering pattern. We also see numerically that the proportion of times a simulation ends in a particular state is determined by the initial state. For example for initial state $|\Psi_0\rangle$ we get the end state in Figure \ref{fig:develop_three_site}(c) with probability $|\langle 2 0 1|\Psi_0\rangle|^2 + |\langle 1 0 2|\Psi_0\rangle|^2$.

%%%%%%%%%%%%%%%%%%%%%%%%%%%%FIGURE%%%%%%%%%%%%%%%%%%%%%%%%%%%%%%%%%%%%%%%%%%%%%%%%%%%%%%%%%%%%%%%%%%%%%%%%%%%%%%%%%%%%%%%%%%%%%%%%%%%%%%%%%%%%%%%%%%%%%%%%%%%%%%%%%%%%%%%%%%%%%%%%%%%%%%%%%%%%%%%%%%%%%%%%%%%%%%%%%%%%%%%%%%%%%%%%%%%%%%%%%%%%%%%%%%%%%%%%%%%%%%

%
%\begin{figure}
%\centering
%\includegraphics{Final_state_and_photon_dist}
%\caption{(a) Proportion of 10000 simulations that end in a final state superposition of (1) $|3 0 0\rangle$, $|0 3  0\rangle$, and $|0 0 3\rangle$, (2) $|2 1 0\rangle$, $|1 2  0\rangle$, $|0 1 2\rangle$ and $|0 2 1\rangle$, (3) $|2 0 1\rangle$, and $|1 0 2\rangle$, and (4) $|1 1 1\rangle$ at various values of $U/J$. The initial probabilities of finding these basis states in the initial ground state as shown by the diamonds. (b) (points) Number of event detections $N_d(\theta)$ after 10000 simulations of 1000 detection events each, where the angular range is divided into 600 bins for event counting. (lines) Scattering distributions predicted from the initial ground states.  Parameters used are $gN=0.5 $ and $k_0 = \pi/a$.}
%\label{fig:state_photon_dist}
%%figure produced by Plot_multiple_sim_end_state_distributions_N.m
%\end{figure}

\begin{figure}
\centering
\includegraphics{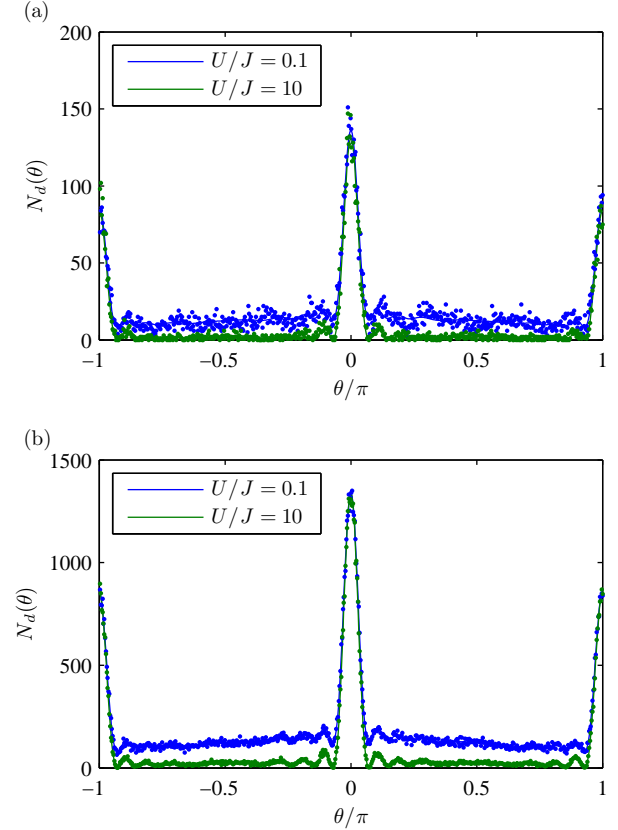}
\caption{Comparison of scattering distribution summed over multiple simulations for a lattice of 9 sites containing 9 atoms for two different ratios of the Bose-Hubbard parameters $U$ and $J$. (points) Number of event detections $N_d(\theta)$ after (a) 10 simulations of 10000 detection events each and (b) 1000 simulations with 1000 detection events each, where the angular range is divided into 600 bins for event counting. (lines) Scattering distributions predicted from the initial ground states.  Parameters used are $g=0.1 $ and $k_0 = \pi/a$.}
\label{fig:Error_in_dist}
%figure produced by Plot_multiple_sim_end_state_distributions_N.m
\end{figure}

%%%%%%%%%%%%%%%%%%%%%%%%%%%%FIGURE%%%%%%%%%%%%%%%%%%%%%%%%%%%%%%%%%%%%%%%%%%%%%%%%%%%%%%%%%%%%%%%%%%%%%%%%%%%%%%%%%%%%%%%%%%%%%%%%%%%%%%%%%%%%%%%%%%%%%%%%%%%%%%%%%%%%%%%%%%%%%%%%%%%%%%%%%%%%%%%%%%%%%%%%%%%%%%%%%%%%%%%%%%%%%%%%%%%%%%%%%%%%%%%%%%%%%%%%%%%%%%

Because the final state proportions are the same as the initial basis state probabilities, we can obtain the initial state scattering distribution if we average over an ensemble of scattering experiments. This occurs even though the initial state is completely changed in the scattering process. For $m$ scattering events per run of the experiment and $n$ runs, each individual run will push the state towards a particular set of basis vectors with a probability determined by the original state. Repetitions of the experiment then lead to a statistical determination of the initial state scattering distribution as we probabilistically sample the scattering distributions for the basis states that make up the initial state. To get the best estimate of the initial state pattern with limited recourses, if we are restricted only by the total number of scattering events $mn$ then the optimal method would be to take $m=1$ in which case back-action does not play a part. In many interesting experiments however, doing an experiment run is more restrictive than scattering. Then the error in estimating the distribution by sampling $n$ runs scales with $1/\sqrt{n}$. Furthermore, when $n\gg 1$ increasing the number of scattering events per run increases accuracy with the same scaling in $m$ and the accuracy of determining the photon distribution scales with $1/\sqrt{mn}$.

In Figure \ref{fig:Error_in_dist} we show how summing over multiple simulations reproduces the initial ground state pattern for a lattice with 9 sites and 9 atoms. Despite there being 6420 unique scattering patterns from the associated basis vectors, sampling only 10 is sufficient to distinguish the two different ground states of the Bose-Hubbard model we have show here. In Figure \ref{fig:Error_in_dist}(a) each of the 10 simulations includes 10000 events of which 10\% (5\%) were scattering events for $U/J = 0.1 (10)$. 
%In fact the results are independent of the scattering rate and coupling constant $g$, where a lower $g$ will simply mean more probe particles are needed to acheive the same number of photons in the scattered pattern. 
In Figure \ref{fig:Error_in_dist}(b) we show the distribution from summing 1000 simulations with 1000 events each, in which case the error in the scattering pattern compared with the exact initial state pattern is reduced by a factor of approximately $\sqrt{10}$ in accordance with the $1/\sqrt{mn}$ scaling.

%%%%%%%%%%%%%%%%%%%%%%%%%%%%%%%%%%%%%%%%%%%%%%%%%%%%%%%%%%%%%%%%%%%%%%%%%%%%%%%%%%%%%%%%%%%%%%%%%%%%%%%%%%%%%%%%%%%%%%%%%%%%%%%%%%%%%%%%%%%%%%%%%%%%%%%%%%%%%%%%%%%%%%%%%%%%%%%%%%%%%%%%%%%%%%%%%%%%%%%%%%%%%%%%%%%%%%%%%%%%%%%%%%%%%%%%%%%%%%%%%%%%%%%%%%%%%%%%

\begin{figure}
\centering
\includegraphics{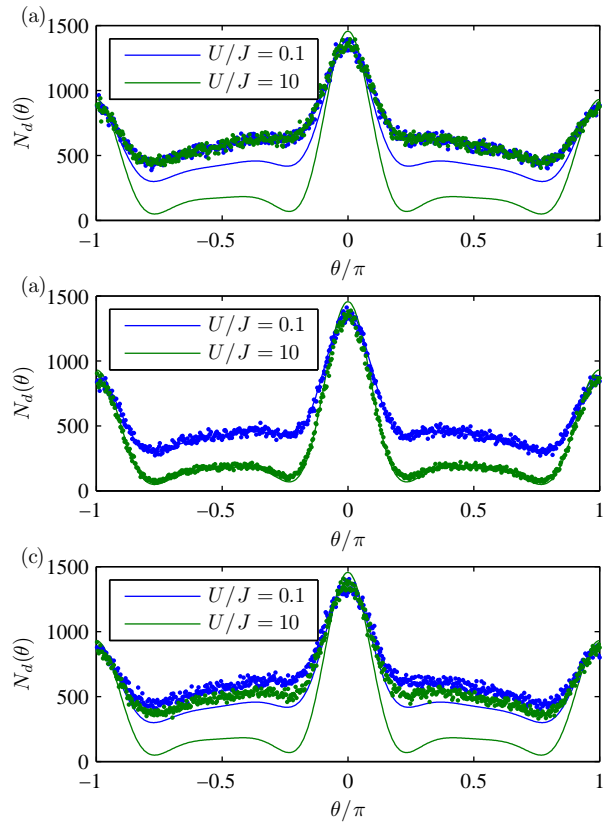}
\caption{Comparison of scattering distributions with different time $dt$ between scattering events for a lattice of 3 sites containing 3 atoms. (points) Number of event detections $N_d(\theta)$ for (a) one simulation of 1000000 events with $dt=1 J/\hbar$, (b-c) 1000 simulations of 1000 detection events each for (b) $dt = 0.001 J/\hbar$, (c) $dt = 0.01 J/\hbar$, where the angular range is divided into 600 bins for event counting. (lines) Scattering distributions predicted from the initial ground states.  Parameters used are $g=0.3 $ and $k_0 = \pi/a$.}
\label{fig:Error_from_evo}
%figure produced by Plot_scattering_dist_for_dif_evo_times.m
\end{figure}

%%%%%%%%%%%%%%%%%%%%%%%%%%%%%%%%%%%%%%%%%%%%%%%%%%%%%%%%%%%%%%%%%%%%%%%%%%%%%%%%%%%%%%%%%%%%%%%%%%%%%%%%%%%%%%%%%%%%%%%%%%%%%%%%%%%%%%%%%%%%%%%%%%%%%%%%%%%%%%%%%%%%%%%%%%%%%%%%%%%%%%%%%%%%%%%%%%%%%%%%%%%%%%%%%%%%%%%%%%%%%%%%%%%%%%%%%%%%%%%%%%%%%%%%%%%%%%%%

The above results applied with no system evolution between scattering events. As we saw in Figure \ref{fig:develop_three_site}(e) system evolution causes the state to eventually progress through the basis states rather than just settling into one group. One might wonder whether the basis states might then be sampled in proportion to the initial state and allow us to get the ground state scattering pattern without having to re-run the experiment multiple times. This is not the case and in fact for simulations long compared to the system's natural evolution time the basis states end up being sampled equally and the scattering pattern is independent of the initial state as shown in Figure \ref{fig:Error_from_evo}(a). There are intermediate regimes however where scattering still typically results in a single group of basis states and the ground state can be recovered by summing over experiments. Figure \ref{fig:Error_from_evo}(b) shows the result of summing over 1000 simulations where each simulation lasted at total time of one natural unit of system time. In this case the summed scattering distribution closely matches the initial state pattern. For the same parameters but a total time of 10 units, Figure \ref{fig:Error_from_evo}(c), we see a significant departure from the initial state pattern and the patterns from the two states begin to merge.

\section{Conclusion}

In conclusion, we have shown how scattering in the presence of measurement backaction scattering leads the initial state of the system to be completely destroyed. When the system does not have time to evolve this leads the system towards eigenstates of the scattering measurement. Each individual experiment then yields a scattering pattern corresponding to a particular eigenstate with probabilities determined by the initial state. Summing these scattering patterns then gives an estimate of the initial state scattering pattern. This allows the initial state to be probed even though it is completely destroyed in each run of the experiment. We have further shown that when the scattering happens over a time that is larger than the natural evolution of system this method no longer yields the initial state pattern, and thus to optimize measurement of the initial state we must scatter all particles within the natural evolution time or less.

%%%%%%%%%%%%%%%%%%%%%%%%%%%%%%%%%%%%%%%%%%%%%%%%%%%%%%%%%%%%%%%%%%%%%%%%%%%%%%%%%%%%%%%%%%%%%%%%%%%%%%%%%%%%%%%%%%%%%%%%%%%%%%%%%%%%%%%%%%%%%%%%%%%%%%%%%%%%%%%%%%%%%%%%%%%%%%%%%%%%%%%%%%%%%%%%%%%%%%%%%%%%%%%%%%%%%%%%%%%%%%%%%%%%%%%%%%%%%%%%%%%%%%%%%%%%%%%%

%\bibliography{../../bib/dphil}

%

\end{document}